\documentclass[11pt,a4paper]{jpconf}
\usepackage[latin1]{inputenc}
\usepackage{setspace}
\usepackage{amsmath}
\usepackage{amsfonts}
\usepackage{amssymb}
\usepackage{graphicx}
\usepackage{wrapfig}

\begin{document}
\title{Half quantum vortices and Majorana fermions in triplet superconductors}

\author{H Y Kee$^1$, A Raghavan$^2$ and K Maki$^2$}
\address{$^1$Department of Physics, University of Toronto, Ontario M5S 1A7, Canada}
\address{$^2$Department of Physics and Astronomy, University of Southern California, Los Angeles CA 90089, USA.}

\begin{abstract}
Half quantum vortices (HQVs) in superfluid $^3He-A$ have been speculated since 1976.  Two years ago, Yamashita et al\cite{yash} reported a very unusual NMR satellite in the rotating superfluid $^3He-A$ in a parallel plate geometry.  Recently this satellite has been interpreted in terms of HQVs\cite{keemaki}.  Also, the strong flux pinnings in the triplet superconductors $UPt_3, Sr_2RuO_4$ and $U_{1-x}Th_xBe_{13}$ have been discovered by Mota and her colleagues\cite{amann}\cite{mota} which are interpreted in terms of HQVs\cite{sigrist}\cite{keekim}\cite{wonetal}.  We shall first review the bound state spectra around a single vortex in unconventional superconductors, which provides the necessary background.  In particular, we find the zero modes of the Majorana fermions in some of triplet superconductors $UPt_3, Sr_2RuO_4$ and $CePt_3Sn$.  Armed with these results we analyze the microscopic aspects of HQVs in $UPt_3$ and $Sr_2RuO_4$.  We compute the quasiparticle density of states and the local magnetic field around HQVs, which should be accessible to scanning tunneling microscopy (STM), micromagnetometry and neutron scattering below 300 mK.
\end{abstract}
\vspace{10mm}

\section{Introduction}	
In this century, unconventional or nodal superconductors take center stage\cite{wonetal2}\cite{makihaas}\cite{wonmosita}. In particular, there are many triplet superconductors like $UPt_3, Sr_2RuO_4, U_{1-x}Th_xBe_{13}$ and $PrOs_4Sb_{12}$, Bechgaard salts $(TMTSF)_2X$ with $X = ClO_4, PF_6, \dots, UNi_2Al_3$ and $CePt_3Si$. Of course, the last superconductor is claimed to be an admixture of singlet and triplet components\cite{wonmosita}. However, the NMR data\cite{yogietal} indicates that the singlet component should be less than 5\%.  Therefore, the superconducting order parameter of $CePt_3Si$ should be similar to the one in superfluid $^3He^{-}A$ and dominated by the single spin component\cite{leggett}\cite{vollhardt}.  Also, the $\sqrt{H}$ dependence of the magnetothermal conductivity in $CePt_3Si$\cite{izawa} at least has the presence of another almost non-superconducting band\cite{makiwon}.  The first triplet superconductor, superfluid $^3He$ was discovered in 1972.  The superfluid $^3He$ appears in three disguises: $^3He-A$, $^3He-A_1$ and $^3He-B$\cite{leggett}\cite{vollhardt}.
\paragraph{}
Most of the superfluid properties of $^3He$ are described in the framework of BCS p-wave superconductors.  Of course, unlike classical s-wave superconductors, these superconductors have large internal degrees of freedom, which manifest themselves as a variety of collective modes and textures\cite{volovik}\cite{monastyrsky}.  We note that the textures in superfluid $^3He-A$ were first introduced by de Gennes\cite{degennes} in analogy to liquid crystals.  Then, Volovik and others\cite{volmin}\cite{cross}\cite{salomaa} have speculated half-quantum vortices (HQVs) in superfluid $^3He-A$.  In the simplest configurations, $\hat{l}$-textures have to be suppressed to have HQVs.  This is realized in the parallel plate geometry where the gap D between 2 parallel plates satisfies the condition $D < 2.3\xi_D\sim23 \mu m$
where $\xi_D(\sim 10\mu m)$ is the dipole coherence length\cite{ambegaokar}\cite{bruinsma}.  In spite of the intensitive search for HQVs by the Helsinki group, HQVs have not been found until recently\cite{hakonen}\cite{volovik2}.  As already mentioned, we believe that Yamashita et al.\cite{yash} have observed HQVs.  The key to their success appears to be the high precision series of parallel plates with $D\simeq10\mu m$, which had not been achieved in earlier experiments\cite{hakonen}\cite{volovik2}.
\paragraph{}
In section 2, we shall first review the bound-state spectra around a simple vortex in unconventional or nodal superconductors.  For s-wave superconductors the result by Caroli, de Gennes and Matricon\cite{caroli}\cite{degennes2} is well known.  More recently, the result is extended to p-wave superconductors, as in superfluid $^3He-A$ and 2D chiral p-wave superconductors\cite{kopnin}\cite{volovik3}\cite{ivanov}\cite{tewari}.  In particular, triplet superconductors have the zero mode with $E=0$ associated with the Majorana fermions\cite{majorana}.  We shall establish that the zero mode exists at least in all triplet superconductors with equal spin pairing (ESP).  For example $UPt_3, Sr_2RuO_4,PrO_4Sb_{12}$ and $CePt_3Si$ belong to ESP.  Then, in section 3 we shall explore HQVs in $UPt_3$ and $Sr_2RuO_4$.

\section{Bound state spectra around a single vortex in unconventional superconductors}
First let us generalize the result of Caroli, de Gennes and Matricon\cite{caroli}\cite{degennes2} for unconventional superconductors.  For many unconventional superconductors we can write\cite{wonetal2}$\Delta(\vec{r},k) = \Delta(\vec{r})f(k)$.  For example, a d-wave superconductor has $f(k) = \cos(2\phi)$.  Also $\Delta(\vec{r})$ around the vortex in a singlet superconductor at $x=y=0$ is well approximated by \cite{gygi}\cite{kato}
\begin{equation}
\Delta(\vec{r}) = \Delta e^{i\phi} th(r/\xi)
\end{equation}
where $\xi \sim v_p/\Delta$ is a variational parameter.
\paragraph{}
Then substituting Eq(1) into the CdGM formula, the bound state spectrum around a vortex is given by
\begin{equation}
\epsilon_n = (n+\frac{1}{2})p_F^{-1}\dfrac{\langle\int_{0}^{\infty}{dr \dfrac{\vert\Delta(r,k)\vert}{r}e^{-2K(r,k)}}\rangle}{\langle\int_{0}^{\infty}{dr e^{-2K(r,k)}}\rangle}
\end{equation}
where,
\begin{align}
K(r,k) = v_F^{-1}\vert f(k)\vert\int_{0}^{\infty}{dr\Delta th(\frac{r}{\xi})}\nonumber\\
=\xi v_F^{-1}\Delta\vert f(k)\vert\ln (\cosh(\frac{r}{\xi}))
\end{align}
Here $\langle\dots\rangle$ means the average over the Fermi surface.  Then we find,
\begin{equation}
\epsilon_n = \omega_0(n+\frac{1}{2})
\end{equation}
with
\begin{eqnarray}
\omega_0 = \Delta(p_F\xi)^{-1}I_1/I_0 \\
I_1 = \frac{1}{2}\langle\int_{0}^{1}{ds(1-s)^{C\vert f\vert-1}[th^{-1}(\sqrt{s})]^{-1}}\rangle \\
I_2 = \dfrac{1}{2}\langle B(\frac{1}{2},C\vert f\vert)\rangle
\end{eqnarray}
where $B(\alpha,\beta)$ is the Eulerian beta function.  Here $C=\Delta\xi/v_F\sim 1$ and $n=0,\pm1,\pm2,
\ldots$.  For s-wave superconductors, Eq.(4) is the CdGM result.  For singlet unconventional superconductors, this gives Volovik's result\cite{volovik3}.  The bound state spectrum is the same as that of the s-wave superconductors, except that
\begin{equation}
\omega_0 = \Delta(p_F\xi)^{-1}\dfrac{\pi}{2}[-\ln(C)]^{-1}\times[\ln(\frac{2}{\delta})]^{-1}
\end{equation}
where $\delta$ is a cut-off parameter for nodal superconductors.  In the presence of impurity scattering, $\delta$ is replaced by $\sqrt{\dfrac{\Gamma}{\Delta}}$, $\Gamma$ being the quasiparticle scattering rate in the normal state.  Therefore, $\omega_0$ is in general a somewhat reduced form of the one for s-wave superconductors.
\paragraph{}
In the chiral triplet superconductors with equal spin pairing (ESP)\cite{kopnin}\cite{ivanov}\cite{tewari}, Eq.(4) is replaced by
\begin{equation}
\epsilon_n = \omega_0n
\end{equation}
with $n=0,\mp1,\mp2,\mp3,\ldots$
\paragraph{}
So there is the zero mode with $\epsilon_0=0$.  The related wave function is obtained following\cite{tewari}
\begin{equation}
\phi_0(r) = (2\pi I_1)^{-1/2}\xi^{-1}(\text{sech}(\frac{r}{\xi}))^{C\vert f\vert}
\end{equation}
where $I_1$ is given by Eq.(6).  So, unlike the case of 2D chiral p-wave superconductors, $\phi_0(r)$ extends to infinity in the nodal directions where $\vert f(k)\vert = 0$.  Also the present result applies for $UPt_3, Sr_2RuO_4, PrOs_4Sb_{12}$ etc.  As already mentioned the zero mode describes Majorana fermions.  Further $\vert \phi_0(r)\vert^2$ is accessible via ultra-low temperature tunneling, with $T<\omega_0\sim 100$ mK.

\section{HQVs in $UPt_3$ and $Sr_2RuO_4$}
First of all the superconducting order parameters in $UPt_3$ and $Sr_2RuO_4$ are described in terms of $\hat{l}$ (chiral vector), $\hat{d}$(spin vector) and $\phi$ the phase of $\Delta(\vec{r})$ similar to the one for superfluid $^3He-A$\cite{sigrist}\cite{keekim}\cite{wonetal}.  Further, $\hat{l}$ is fixed parallel to the crystalline $\overrightarrow{c}$-axis.  Also, in the absence of magnetic fields, $\hat{d}\Vert\hat{l}$ just like in superfluid $^3He-A$.  Of course, the relevant energy for $\hat{d}\Vert\hat{l}$ comes from the spin-orbit term unlike superfluid $^3He-A$.  Nevertheless, we use $\xi_D\sim\mu m$ in order to describe the relevant length scale.
\paragraph{}
When a magnetic field $\overrightarrow{H}\Vert\overrightarrow{c}$ is applied, the uniform texture $\hat{d}\Vert\hat{l}$ is broken in a magnetic field close to the upper-critical field $H_{c_2}(t)$.  We denote the relevant length scale by $\xi_H$.  Then the texture free energy takes the same form as in the superfluid $^3He-A$\cite{keekim}
\begin{equation}
F=\frac{1}{2}\chi_NC^2\int dxdy\lbrace K(\bigtriangledown\phi)^2+\sum_{i,j}\vert\partial_i\hat{d}_j\vert^2+\xi_D^{'-2}(\hat{d}_x^2+\hat{d}_y^2)\rbrace
\end{equation}

where $\chi_N$ and $C$ are the spin susceptibility and the spin wave velocity and $\xi_D^{'-2}=\xi_D^{-2}-\xi_H^{-2}$.  Here
\begin{equation}
K = \dfrac{\rho_s(t)}{\rho_{sp}(t)}=\frac{1+\frac{1}{3}F_1^a}{1+\frac{1}{3}F_1}.\frac{1+\frac{1}{3}F_1^a(1-\rho_s^0(t))}{1+\frac{1}{3}F_1(1-\rho_s^0(t))}
\end{equation}
and $\rho_s(t)$, $\rho_{sp}(t)$ and $\rho_s^0(t)$ are the superfluid density, the spin superfluid density and the bare superfluid density; $F_1$, $F_1^a$ are the Landau parameters.  Further, $\frac{m^*}{m}=1+\frac{1}{3}F_1$ are known for $UPt_3$ and $Sr_2RuO_4$ ($\gamma$-band) as 20 and 16 respectively.  Therefore, if we neglect $F_1^a$ in $UPt_3$ and $Sr_2RuO_4$ we can evaluate $K(t)$ as shown in Fig.(1).  In calculating $K(t)$, we need $\rho_s^0(t)$ for $E_{2u}$ and the chiral f-wave superconductors, which is given by the one for d-wave superconductors\cite{deguchi} in the weak coupling limit:
\begin{figure}
\begin{center}
\includegraphics[width=4.0in]{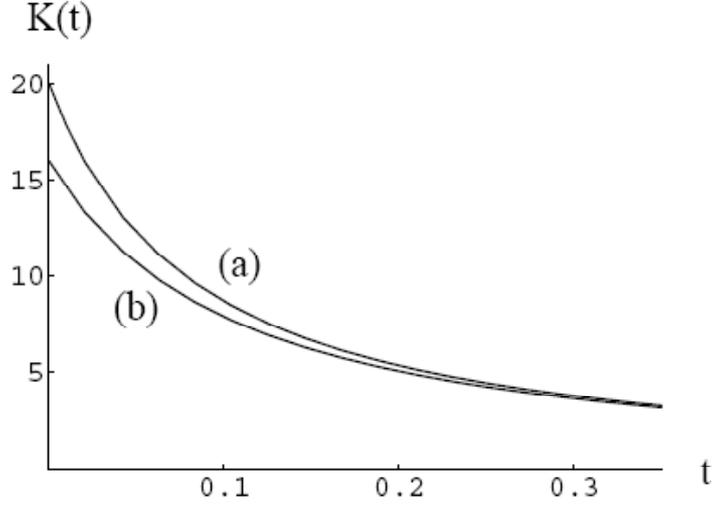}
\end{center}
\caption{$K(t)$ is shown for (a)$UPt_3$ and (b)$Sr_2RuO_4$}\label{fig1}
\end{figure}
\begin{equation}
\rho_s^0(t) = 1-0.647 t - 0.335 t^2 + 0.02 t^3
\end{equation}
Now let us consider the stability region of HQVs in a magnetic field $H$ parallel to the $c$-axis.  For this purpose it is convenient to divide the 2D plane (i.e. the u-b plane) in circles of radius a, which encloses a single flux of quantum $\phi_0 = 2.07\times10^{-7} Tcm^2$.  This gives $a=(\frac{\phi_0}{\pi H})^{\frac{1}{2}}$.  Noting that the upper critical field is given by
\begin{eqnarray}
H_{c_2}(t) = \frac{\phi_0}{2\pi\xi^2(t)}\nonumber\\
a=\sqrt{2}\xi(t)
\end{eqnarray}
at $H=H_{c_2}(t)$.
\paragraph{}
Making use of Eq.(11), the free energy of Abrikosov's vortex in a unit cell is given by
\begin{equation}
F_A = \pi\chi_NC^2\ln(\frac{a}{\xi})
\end{equation}
On the other hand, a pair of HQVs in a unit cell is given by
\begin{eqnarray}
F_{BP} = \frac{\pi}{2}\chi_NC^2\lbrace K\ln\left(\dfrac{a+\sqrt{a^2-\frac{R^2}{4}}}{2\xi}\right)+\ln(\frac{a}{\xi})\nonumber\\
-\frac{R}{4a}\sin^{-1}\left(\frac{R}{\sqrt{a^2-\frac{R^2}{4}}}\right)+\frac{1}{4}(\frac{R}{\xi'_D})^2\ln(\frac{4\xi'_D}{R})\rbrace
\end{eqnarray}
where $R\leq 2a$ is the distance between the pair.  For $K\gg1$, R is well approximated\cite{keemaki} by $\frac{R}{2a}=\frac{\sqrt{2K+1}}{K+1}$.  Then Eq.(16) reduces to
\begin{eqnarray}
F_{BP} = \frac{\pi}{2}\chi_NC^2\lbrace K\ln\left(\dfrac{a}{2\xi}\dfrac{2K+1}{K+1}{2\xi}\right)+\ln(\frac{a}{\xi})
-\frac{1}{2}\frac{\sqrt{2K+1}}{K+1}\sin^{-1}\left(\frac{2\sqrt{2K+1}}{K+1}\right)\nonumber\\
+\frac{2K+1}{(K+1)^2}(\frac{a}{\xi'_D})\ln(\frac{2\xi'_D}{a}\frac{K+1}{\sqrt{2K+1}})\rbrace
\end{eqnarray}
Finally, equating $F_A = F_{BP}$ gives,
\begin{equation}
\frac{a}{\xi}=\left(2(2K+1)^{\frac{1}{2}(2K+1)}(K+1)^{-(k+1)}\right)^{\frac{1}{K-1}}.\left(\frac{K+1}{\sqrt{2K+1}}\frac{\xi'_D}{a}\right)^{\frac{2K+1}{(K+1)^2(K-1)}(\frac{a}{\xi'_D})^2}
\end{equation}
In particular, assuming that $\xi_H=\xi_D$ at $H=H*(t)$, we find
\begin{equation}
\frac{H^*(t)}{H_{c_2}} = \left(2^{-(K-2)}(2K+1)^{(2K+1)}(K+1)^{-2(K+1)}\right)^{-\frac{1}{K-1}}
\end{equation}
Now making use of the $K(t)$'s shown in Fig.(1), $H^*(t)/H_{c_2}(t)$ for $UPt_3$ and $Sr_2RuO_4$ are obtained as shown in Fig.(2).  Surprisingly, $H^*(t)/H_{c_2}(t)$ for $UPt_3$ and $Sr_2RuO_4$ are not distinguishable.  We may speculate that $H^*(t)$ corresponds to the B-C boundary in $UPt_3$.  No such boundary has been established for $Sr_2RuO_4$.  However, considering the anomalous magneto-specific heat data for $Sr_2RuO_4$ below 300 mK\cite{deguchi}\cite{katomaki}, it is possible that the extra angular dependence of the magneto-specific heat is due to the Abrikosov vortex to HQVs transformation.  Clearly, further studies of HQVs in $Sr_2RuO_4$ is highly desirable.
\begin{figure}
\begin{center}
\includegraphics[width=3.0in]{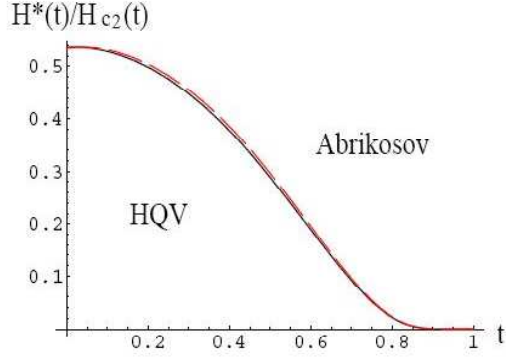}
\end{center}
\caption{$H^*(t)/H_{c_2}(t)$ is plotted as a function of $t$.  The red-line is for $UPt_3$}\label{fig2}
\end{figure}
\paragraph{}
When $\hat{d}\bot\hat{l}$ everywhere, Ivanov\cite{ivanov} has shown that the spin up and spin down components are decoupled in the Bogoliubov de Gennes equation.  Then one of the HQV pairs consists of a simple quantum vortex with, say, spin up, while the other is the one with spin down.
\paragraph{}
In this particular case, we can write down the quasiparticle density of states as
\begin{equation}
N_{\text{pair}}=\frac{1}{2}\left(N(\overrightarrow{r}+\frac{\overrightarrow{R}}{2},E)+N(\overrightarrow{r}-\frac{\overrightarrow{R}}{2},E)\right)
\end{equation}
where $\overrightarrow{R}=R\hat{x}$ and $N(\overrightarrow{r},E)$ is the quasiparticle density of states associated with a single Abrikosov vortex\cite{katomaki}.  Similarly the local magnetic field is given by
\begin{equation}
\overrightarrow{H}_{\text{pair}}=\frac{\phi_0\hat{z}}{4\pi\lambda}\left(K_0\left(\frac{\vert\overrightarrow{r}+\frac{\overrightarrow{R}}{2}\vert}{\lambda}\right)+K_0\left(\frac{\vert\overrightarrow{r}-\frac{\overrightarrow{R}}{2}\vert}{\lambda}\right)\right)
\end{equation}

\begin{figure}
\begin{center}
\includegraphics[width=4.0in]{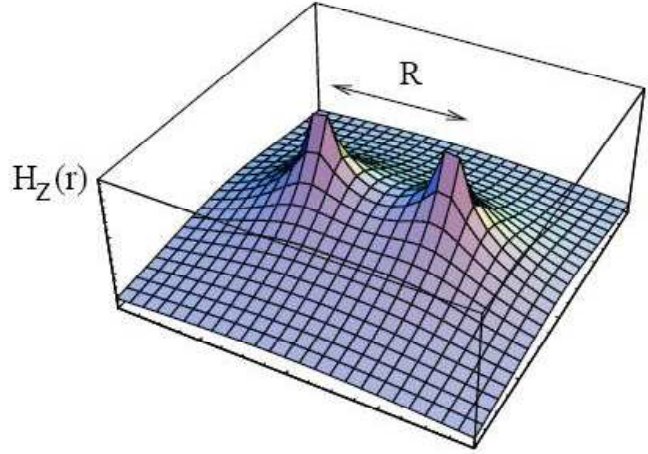}
\end{center}
\caption{The local magnetic field generated by a pair of HQVs}\label{fig3}
\end{figure}
where $K_0(z)$ is the mollified Bessel function\cite{volmin} and $\lambda$ is the magnetic penetration depth.  We have shown a sketch of Eq.(21) in Fig.(3).  Also, each spin component has the zero mode associated Majorana fermion.  In reality, however, the condition $\hat{d}\bot\hat{l}$ is most likely broken near the centers of HQVs.  We believe Eq.(20) is still valid in the circumstance while the length scale and the energy scale in Eq.(20) have to be changed accordingly.  Nevertheless, we believe both Eq.(20) and Eq.(21) provide a good guide for experimentalists.  These are surely accessible via STM (scanning tunneling microscopy)\cite{maggio}\cite{fischer}, micromagnetometry\cite{kirtley}\cite{tsuei} and neutron scattering experiments below 300 mK.

\ack
We have benefited from discussions with Balazs Dora, Stephan Haas, Corneliu Miclea, Peter Thalmeier and Grisha Volovik.  KM acknowledges the hospitality of the Max-Planck institute for the \textit{Physics of Complex Systems} at Dresden, where a part of this work was done.

\end{document}